# Substrate Effect on Plasmonic Cube Dimers: Reversal of Optical Binding Force Induced by Strong Fano Resonance

M.R.C. Mahdy*[1,2], Tianhang Zhang[1,3], Md. Danesh[1], Weiqiang Ding*[4]

[1]Department of Electrical and Computer Engineering, National University of Singapore, 4 Engineering Drive 3, Singapore 117583

[2]Department of Electrical & Computer Engineering, North South University, Bashundhara, Dhaka 1229, Bangladesh

[3]NUS Graduate School for Integrative Sciences and Engineering, National University of Singapore, 28 Medical Drive, Singapore 117456

[4]Department of Physics, Harbin Institute of Technology, Harbin 150001, People's Republic of China

* Corresponding authors: mahdy.chowdhury@northsouth.edu and wqding@hit.edu.cn

**The behavior of Fano resonance and the reversal of near field optical binding force of dimers over different substrates have not been studied so far. In this work, we observe that if the closely located plasmonic cube homodimers over glass or high permittivity dielectric substrate are illuminated with plane wave polarized parallel to dimer axis, no reversal of optical binding force occurs. But if we apply the same set-up over a plasmonic substrate, stable Fano resonance occurs along with the reversal of near field binding force. It is observed that during such Fano resonance stronger coupling occurs between the dimers and plasmonic substrate along with the strong enhancement of the substrate current. Such near field binding force reversals of plasmonic cube dimers have been explained based on the observed unusual behavior of optical Lorentz force during the induced stronger Fano resonance and the dipole-dipole resonance. Although previously reported reversals of near field optical binding forces were highly sensitive to particle size/shape (i.e. for heterodimers) and inter-particle distance, our configuration provides much relaxation of those parameters and hence should be verified experimentally with simpler experimental set-ups.**

**Keywords:** Fano resonance; Optical force; Plasmonic substrate; Binding force; Plasmonic dimer; Lorentz force

**INTRODUCTION**

Since its discovery, Fano resonance has been a typical feature of interacting quantum systems. However, recently the Fano resonance has been found in plasmonic nanoparticles, photonic crystals, and electromagnetic metamaterials. Fano resonance and the Fano line shape [1] cannot be described by the conventional and well-known Lorentz formula [2]. In plasmonic nanostructures, Fano resonance can happen due to the resonant destructive interference between a super-radiant (bright) mode and subradiant (dark) modes [3]. The promising applications of plasmonic hybridization and Fano resonances [4,5] have been investigated in improved sensitivity of the resonance [6], bio sensing [7], surface-enhanced Raman scattering [8], photonic propagation and wave guiding [9], plasmon-induced transparency [10] and many others [11]. In contrast, much less attention is dedicated on near field optical force due to Fano resonance; especially for plasmonic dimers [4,5] as discussed only in [12,13]. Specially, plasmonic particle over substrate show remarkable properties: Fano resonances [14], broadband tenability in plasmonic resonance [15], modification of energies of the plasmon modes [16], changing the localized density of states [17], radiative enhancement [18], frequency shift of an electric dipole resonance [19] and so on. But the behavior of near field optical force for plasmonic dimers over different types of substrate and the effect of substrate on the reversal of near filed optical binding force have not been studied in literature. In fact, only few works have been reported on the reversal of near field optical binding force due to Fano resonance, i.e. for nanobar structures [12], for disk along with a ring structure [13] and those are highly dependent on inter-particle distance and particle size.

According to ref. [20]: "The inter particle separations are usually comparable to the wavelength of the illuminating laser beam and, therefore, mainly long-range interactions mediated by the far-field scattered field are considered while the near-field coupling is usually omitted in the related studies." But with the recent technology, the inter particle distance between the nano particles can be precisely controlled [21]. More importantly, the size and shape of the gap between the nanoparticle and film can also be controlled to sub nanometer precision bottom-up fabrication approaches [21]. The film-coupled nanoparticle geometry has recently been applied to enhancing optical fields, accessing the quantum regime of plasmonics [22], and the design of surfaces with controlled reflectance [23]. Still no investigation has been carried out regarding the behavior of both Fano resonance and the reversal of near field optical binding force with respect to the inter particle distance of two dimers over different substrates.

In general, reversal of near field lateral optical binding force for the same polarization of light is quite uncommon with the homodimers placed without substrates [24-27]. Even if the homodimers are placed over substrates, reversal of near field lateral optical binding force has not been observed in refs. [28-30] for spherical shaped and rod shaped plasmonic homodimers. On the other hand, it is well known that Fano resonance is ultra-sensitive and induced reversal of optical binding force dies out very quickly if the inter particle gap of the heterodimers [12] increase even a little bit. Moreover, this Fano resonance induced lateral binding force reversal is highly dependent on particle size [12], which has not been achieved by homodimers so far. In this work, it is observed that if two plasmonic cube homodimers (also applicable for heterodimers) are placed closely without any substrate or above glass or high refractive index substrate, no reversal of optical binding force occurs. But if we apply the same set-up over a plasmonic substrate, stable and stronger Fano resonance occurs along with the reversal of lateral binding force. Importantly, the inter-particle distance and the sizes of the cube homo-dimers are much less sensitive than the previously reported case in [12].

In this article, we have investigated a fully alternative way in comparison with refs [12, 28-30] to achieve strong and stable reversal of optical lateral binding force based on cube homodimer and substrate system, where stable and much stronger Fano resonance can be achieved due to the substrate coupling with larger area of the lower side of cube dimers. Though nanoparticles of sphere are easy to synthesize in experiment, the substrate-induced hybridization of the plasmon modes can be much larger [14] for a planar metallic nanoparticle (i.e. a cube) deposited on a plasmonic substrate [21,23] than for a spherical particle [16]. This happens because the plasmon-induced steady state currents and surface charges will be located closer to the strong surface currents and plasmonic screening charges induced on the surface of the substrate respectively over a large area.

We have demonstrated that from the interplay between localized surface plasmon and propagating surface plasmon polariton along with the strong coupling between the two particles with the plasmonic substrate, the reversal of the optical binding force occurs during the strong Fano resonance. According to several full wave numerical results, our configuration provides much relaxation on some critical parameters (i.e. inter particle distance and particle size as stated previously) and hence can probably be verified experimentally with much simpler experimental set-ups.

**RESULTS AND DISCUSSIONS**

We specify that throughout this paper we refer to 'exterior' or 'outside' forces as those evaluated outside the volume of the macroscopic particles, while 'interior' or 'inside' refer to those quantities inside this object volume. In order to consider the realistic effects, we have done all the numerical calculations using full wave simulations [31] in three dimensional (3D) structures.

The proposed simple set-up is illustrated in Fig. 1(a). The Silver particles are placed near to each other above a silver substrate (whole set-up is embedded in water). Spacer (refractive index 1.4) of height 5 nm is always considered above all the substrates in this article. The real and imaginary part of the permittivity of silver is taken from the standard Palik data [31, 32]. Inter particle distance is '$d$'. The source is a simple $x$-polarized plane wave $E_x = E_0 e^{-i\beta z}$ propagating towards $-z$ direction and $E_0$ has been set to 1V/m (so, the incident intensity is 0.001176 W/m$^2$) in the simulation model. Our calculated force value is in the range of $10^{-24}$ Newton. On the other hand, in [27] the incident intensity is $0.05 \times 10^{12}$ W/m$^2$ and their force is in Pico-Newton range. So, force in our work would also appear in Pico-Newton range ($10^{-24}/0.001176 \times 0.05 \times 10^{12} \sim 42.52$ pN), if the incident intensity in our work were $0.05 \times 10^{12}$ W/m$^2$ instead of 0.001176 W/m$^2$. Our set-up is a symmetry broken system which later plays a vital role for the force reversal. Details on the calculation process of optical force (based on both Minkowski stress tensor and modified Lorentz force) has been discussed in the 'METHOD' section (at the end of this article).

**Plasmonic cubes over plasmonic substrate: Resonance modes and the reversal of binding force**

In Fig. 1(a) we have considered two silver cubes of length 120 nm each (100 nm apart). These cubes are placed in water above a silver substrate. A spacer of 5 nm height is considered between the cubes and the substrate and the cubes are placed 5 nm above the spacer. This set-up is shined by a plane wave described in previous section. The extinction spectra in Fig. 1(b) reveals that Fano dip takes place at around 754 nm and the strength of extinction coefficient increases two times than a single cube placed over silver substrate. The quadropolar resonance (at wavelength 627 nm), dipolar resonance (at 966 nm) and Fano dip (at around 754 nm) do not experience shifting due to the addition of the identical cube.

It is demonstrated in Fig. 1(c) that reversal of optical binding force [i.e. positive value of $F_{Bind\ (x)}$ means attractive force and negative value means repulsive force] takes place near the Fano dip at 754 nm wavelength. It should also be noted that if we decrease the inter particle distance of the cubes, the magnitude of the lateral binding force ($F_{Bind\ (x)} = (F_{1\ (x)} - F_{2\ (x)})$) increases. Here subscript (x), (1) and (2) represent: +x direction, left cube and right cube respectively.

The distinct charge distributions of the plasmonic cubes have been shown in Figs 1(d) – 1(f) for different resonance modes [i.e. dipole-dipole (DD), Fano and quadrupole-quadrupole (QQ) resonance respectively]. In addition, in Figs 1(g)-(i) the electric field magnitudes of different modes [i.e. DD, Fano and QQ resonance respectively] have been demonstrated. The field distributions (field patterns) look similar (but certainly not identical) to ref. [14] where the substrate is considered as high refractive index dielectric.

At the DD resonance mode (i.e. 966 nm wavelength), field lines form closed loops around the separation between the particle and the metallic film [33]. Maximum value of the repulsive binding force occurs at this wavelength as shown in Fig. 1(c). This resonance mode corresponds to the localized surface plasmon (LSP) mode associated to the silver particle. According to another point of view [21], this resonance mode can be explained as a magnetic cavity mode that roughly satisfies the half wavelength criteria. In particular, the metal (silver)−insulator (spacer)−metal (silver) region supports a transmission line type of mode modified by the plasmonic dispersion of the metal. A large effective impedance mismatch occurs at the edges of the cube and most of the energy is reflected back under the cube [cf. the electric field distribution in Fig. 1(g)].

On the other hand, a resonance peak of extinction spectra occurs at around 627 nm as shown in Fig. 1 (c), which corresponds to QQ resonance. Local maxima of the attractive binding force occur during this resonance. It corresponds to the surface plasmon polariton (SPP) mode propagating on the silver/air interface of the metallic film [cf. ref. [33]], which is excited near this wavelength according to ref. [33]. The corresponding enhancement of the particle excitation field increases the intensity of field scattered at infinity [cf. the electric field distribution in Fig. 1(i)]. From another point of view [21], this higher order mode is an example of waveguide modes and occurs at shorter wavelengths such that the cavity is longer than a half wavelength of the mode.

**Plasmonic cubes over plasmonic substrate: Further discussion on Fano resonance mode**

In supplement S1 it is shown that the reversal of binding force still occurs at the Fano dip position (around 1000 nm) even if we change the length of the cube from 120 to 180 nm. Such reversal of binding force is quite generic for other sized plasmonic cubes over plasmonic substrates. The only countable issue to achieve such force reversal is to achieve Fano dip at first with the single cube. However, the force reversal wavelength and the Fano resonance both red shift with the increase of the cube dimension.

In our optical set-up, the reason of Fano resonance for the dimers can be very easily explained based on the destructive interference of quadropolar and dipolar mode of the dimer as previously discussed considering high refractive index substrate in ref. [14]. The strength of electric field magnitude increases a lot (in comparison with all other substrates) due to the existence of plasmonic substrate, which is a necessary condition for stronger coupling and resonance that leads to binding force reversal. However, the mixed nature of electric field at the Fano dip clearly supports the mixing of dipolar and quadrupolar resonance [14, 15].

But the question is whether Fano resonance is the only factor to achieve such force reversal. The answer is: enhanced Fano resonance induced by strong plasmonic coupling is the key factor of such force reversal but it is not the only reason. For example- in the next section we shall demonstrate that Fano resonance also occurs for Silicon substrate with 120 nm lengthen cube dimers but no reversal of binding force takes place for such case. In addition, more details on the physical mechanism of such optical force reversal will be discussed in the next sections.

**Plasmonic cubes over different substrates: effect on binding force**

So far we have discussed regarding single or double plasmonic cubes over plasmonic substrates. Now we shall compare the behavior of optical binding force considering two plasmonic cubes without substrate and over the glass or Silicon (high refractive index) or silver (plasmonic substrate) embedded in water. In Fig. 2 (a) –(c), no reversal of optical binding force is observed in Fig. 2 (c). This same conclusion is true for the $2^{nd}$ case [cf. Figs. 2(d)-(f)] when no reversal of optical binding force is observed. Albeit, in Fig. 2(g)-(i), the substrate is Silicon (modelled as Palik data [31, 32]), Fano dip spectral region is observed in Fig. 2(h) due to Fano resonance. But no reversal of optical binding force is observed in Fig. 2 (i). This is an important observation. This observation suggests that Fano resonance is not the only criteria to achieve the reversal of optical binding force [which occurs for silver slab in Fig. 2 (l)] rather some other factors (will be discussed next) significantly influence the origin of the reversal of optical binding force. Now we shall discuss the behavior of binding force distribution for these three resonances: (i) DD (ii) QQ and (iii) Fano resonance, when the substrate is silver one.

For the DD resonance mode, the resonance process with the plasmonic substrate is quite different in comparison with the other substrates. During the DD resonance mode, the accumulated charges are strongly coupled with the plasmonic substrate and the total power is mostly concentrated between the cube and the plasmonic substrate. The charge distribution essentially results in a current loop between the nano cube and metal film inducing a strongly enhanced magnetic field. This resonance is then mainly of magnetic nature [21, 23, 33]. But the most interesting part is that: though the resonance is of magnetic nature, the difference of the scattering part [34] (which originates from magnetic induction: $\boldsymbol{J} \times \boldsymbol{B}_{in}$; where $\boldsymbol{J}$ is the polarization induced current [34]) or bulk part of the total Lorentz force [cf. Eq. (4) in the 'METHOD' section] vanishes during the peak of LSP or DD resonance mode [cf. supplement S2 Fig. 2s (a), (c) for both 120 and 180 nm lengthened cubes]:

$$\text{Del } F_{\text{Bulk}(x)} = \int [\langle \boldsymbol{f}_{\text{Bulk}(1)} \rangle dv_{(1)}] - \int [\langle \boldsymbol{f}_{\text{Bulk}(2)} \rangle dv_{(2)}] \qquad (1)$$

Here $\langle \boldsymbol{f}_{\text{Bulk}(j)} \rangle \simeq -\frac{1}{2} \text{Re}[i\omega(\varepsilon_{s(j)} - \varepsilon_b)\{\boldsymbol{E}_{\text{in}(j)} \times \boldsymbol{B}^*_{\text{in}(j)}\} + i\omega(\mu_{s(j)} - \mu_b)\{\boldsymbol{D}^*_{\text{in}(j)} \times \boldsymbol{H}_{\text{in}(j)}\}]$ and j (=1 or 2) represents the left (denoted as '1') or right cube (denoted as '2') respectively in the dimer set. So, the total binding force at LSP or DD resonance peak is fully due to the surface charges and the difference of surface force [cf. Eq. (3) in the 'METHOD' section] can be expressed as:

$$\text{Del } F_{\text{Surf}(x)} = \int [\langle \boldsymbol{f}_{\text{Surface}(1)} \rangle ds_{(1)}] - \int [\langle \boldsymbol{f}_{\text{Surface}(2)} \rangle ds_{(2)}] \qquad (2)$$

Here

$$\langle \boldsymbol{f}_{\text{Surface}(j)} \rangle = \{\epsilon_b (\boldsymbol{E}_{\text{out}(j)} - \boldsymbol{E}_{\text{in}(j)}) \cdot \hat{\boldsymbol{n}}\} \left( \frac{\boldsymbol{E}_{\text{out}(j)} + \boldsymbol{E}_{\text{in}(j)}}{2} \right)^*_{\text{at } r=a} + \{\mu_b (\boldsymbol{H}_{\text{out}(j)} - \boldsymbol{H}_{\text{in}(j)}) \cdot \hat{\boldsymbol{n}}\} \left( \frac{\boldsymbol{H}_{\text{out}(j)} + \boldsymbol{H}_{\text{in}(j)}}{2} \right)^*_{\text{at } r=a}.$$

It should be noted that $F_{\text{Bind}(x)} = (F_{1(x)} - F_{2(x)}) = \text{Del } F_{\text{Bulk}(x)} + \text{Del } F_{\text{Surf}(x)}$. In Fig. 3(i) it can be seen that the flipped $\int [\langle \boldsymbol{f}_{\text{Surface}(1)} \rangle ds_{(1)}]$ [and also Del $F_{\text{Surf}(x)}$; cf. Fig 2s (b) and (d) in supplement S2] reaches its maximum negative value near the DD resonance for plasmonic silver substrate [also cf. Fig 2(l)]. From the first row of Fig. 3, it can also be seen that such reversal of $\int [\langle \boldsymbol{f}_{\text{Surface}(1)} \rangle ds_{(1)}]$ does not happen for other substrates. This can be explained based on the electric field coupling between the cubes and also between cubes with plasmonic substrate. Electric field enhancement becomes much stronger for the case of plasmonic substrate in comparison with other substrates especially at this particular resonance. The surface force in Eq. (3) in the 'METHOD' section depends on the electric fields of the cube boundary/interface, which becomes much stronger during this DD resonance.

Though the maximum value of repulsive binding force for DD or LSP mode is just due to surface charges, the local maximum value of attractive binding force for QQ or SPP resonance at wavelength 627 nm [cf. Fig 2(l)] is the combined effect of static (opposite surface charges as shown in Fig. 1 (f)) and dynamic (enhanced/stronger propagation of SPP as shown in supplement S3) process. If we give a close look at Lorentz force in supplement S2, it is clearly observable that at this QQ resonance: though the resultant binding force is due to the dominance of Del $F_{Surf(x)}$, Del $F_{Bulk(x)}$ also exists. This fact can be verified from Fig. 3(i), where the dominance of $\int [\langle f_{Surface(B)} \rangle ds_{(1)}]$ on cube-1 is clearly observable. The opposite charges of the cubes create the attractive force between the cubes placed over any substrate such as plasmonic or glass or silicon substrate. At the same time, the propagating surface plasmon of the plasmonic substrate also create force on the cubes in the opposite direction of its propagation, which ultimately causes the local maximum value of attractive force at this QQ or SPP resonance [cf. Fig 2(l) in main article and Fig 3s in supplement S3]. On the other hand, the strength of SPP becomes extremely weak at DD or LSP resonance as shown in supplement S3.

Now we shall consider the behavior of optical force for Fano resonance when the reversal of lateral binding force takes place [cf. Fig 2(l)]. If we compare the extinction efficiency of the four cases [cf. the second column of Fig 2], the magnitude is much stronger for the plasmonic (silver) substrate case in comparison with all other cases. In addition, if we consider the strength of induced current at resonant frequencies (i.e. Fano resonance and previously discussed DD resonance mode), the surface current for the case of silver slab is quite stronger in comparison with all other cases: no substrate, glass substrate and silicon substrate case. Now if we give a closer look to the individual cube (i.e. cube-1): it is clearly observable that the bulk force part [cf. Eq. (4) in 'METHOD' section] turns into negative force only near the Fano resonance regime for the case of plasmonic substrate. However, this bulk force (which is mainly connected with the scattering force of plasmonic objects [34, 35]) is never dominant, when the substrate is glass or silicon. The total force of cube-1 for glass and silicon substrate case is always dominated by surface force. In contrast, the scenario is fully reversed for the case of silver substrate especially during the Fano resonance. In Fig. 3(i) it can be observed that the reversal of optical binding force occurs due to the reversal of bulk Lorentz force. Our ultimate conclusion is that: (i) the reversal starts to occur from strong multiple scattering and (ii) though it is commonly believed that internal wave-field does not contribute much for optical force on plasmonic objects (cf. the dominance of surface force in [35, 36]), during the Fano resonance the scenario is quite different.

**Effect of height, size and background material on binding force**

In previous section we have discussed the effect of different substrates on lateral binding force. In this section we shall focus our discussion on lateral binding force due to the change of dimer heights from the plasmonic substrate.

In Fig. 4(a)-(f), it is shown that when the cubic silver dimers are far away (45 nm away from the spacer) from the plasmonic substrate, Fano resonance does not occur. For these dimers, placed 45 nm away from the spacer, the dominant resonance in Fig. 4(a) is the DD resonance [but for the isolated dimer case in Fig. 2(a) the dominant resonance is not DD mode [14]]. It should be noted that the binding force at DD resonance mode behaves quite differently for these two cases: (i) Far from the substrate- strong mutual attractive force occurs and (ii) very close [i.e. 5nm away from spacer as shown in Fig. 2 (j)-(l)] to the substrate- strong mutual repulsive force occurs [the reversal mainly occurs before this dipole resonance at the strong Fano resonance, when stronger coupling with the plasmonic substrate starts to occur].

Now, in Fig. 4(g)-(l), when the cubic silver dimers are closely placed (15 nm away from the spacer) from the plasmonic (silver) substrate, Fano resonance does not die out. However, such Fano resonance is not strong enough and optical binding force does not reverse. This configuration can better be comparable with the case: plasmonic homodimers (cube) over the previously discussed Si substrate. The current, electric fields and magnetic fields of such set-ups are also similar (but not identical) to that previously discussed case of dimers over Si substrate. Now, in Fig. 4 (m)-(r), when the cubic silver dimers are closely placed (10 nm away from the spacer) from the plasmonic substrate, strong Fano resonance just starts to occur along with strong coupling with the substrate. During this resonance, the bulk part of Lorentz force reverses and just starts to dominate the total force on each dimer [cf. Fig. 4(o)]. Ultimately these lead to the reversal of optical binding force as shown in Fig. 4 (n). The mechanism of binding force for this case is very similar to the case [5 nm away from the substrate] discussed previously in details in previous section.

The waveguide description [37] of resonance mode [21, 23] also aids in understanding the observed dependence of the resonance shift on both the size of the nano cube [as can be seen from the comparison of Fig. 1(b), (c) and supplement S1 Fig 1s (b), (c)] and the gap between the silver cube and the silver substrate [as shown in Figs. 4]. (1) For larger cube, the length of the cavity increases and will therefore support a resonance at a longer wavelength. As a result, the resonance redshifts. (2) Similarly, when the gap between the substrate and the cube gets smaller, the effective refractive index of the cavity mode increases. This effectively lengthens the cavity [21] and results in a resonance condition at a longer wavelength. For case (1) and (2), when the second cube is placed closely to the first cube, the induced stable Fano resonance also red shifts following the rapid red shift of DD resonance and hence the binding force reversal wavelength also red shifts following the Fano dip in Fig. 4 and Figs 1 (b), (c); supplement S1 Fig 1s (b), (c).

Throughout the article, the background material has been considered water. The simplest way to blue shift all the resonances along with the wavelength of binding force reversal is to decrease the background refractive index. Last but not least, with the increase of the inter-particle gap, the binding force magnitude decreases slowly but still this overall set-up provides a very relaxed mechanism to verify the reversal of optical binding force.

**CONCLUSIONS**

In summary, we have investigated a very simple possible configuration to demonstrate the reversal of lateral optical binding force with plasmonic homodimers based on strong Fano resonance. Among all the substrates (i.e. glass, Si, Ag, Au etc.), the closely placed plasmonic particles should remain very close only to the plasmonic substrate so that the bulk part of the total Lorentz force dominates the total lateral force of each cube during the substrate mediated Fano resonance. The surface current (along with the strong electric and magnetic coupling) increases significantly only for the close presence of the plasmonic substrate, which bears significant influence on the reversal of binding force of the dimers. Though we have shown only the homodimer cases, our proposed idea should also work for the cube heterodimers providing more flexibility on particle size. We believe that our proposals can be verified experimentally due to the simplicity of the proposed set-ups; and thus the attractive and repulsive forces between two plasmonic objects can be robustly adjusted based on the idea proposed in this work.

**METHODS**

The 'outside optical force' [38] is calculated by the integration of time averaged Minkowski [38, 39] stress tensor at r=$a^+$ employing the background fields of the scatterer of radius $a$:

$$\left\langle \boldsymbol{F}_{\text{Total}}^{\text{Out}} \right\rangle = \oint \left\langle \overline{\overline{\boldsymbol{T}}}^{\text{out}} \right\rangle \cdot d\boldsymbol{s}$$
$$\left\langle \overline{\overline{\boldsymbol{T}}}^{\text{out}} \right\rangle = \frac{1}{2}\text{Re}[\boldsymbol{D}_{\text{out}}\boldsymbol{E}_{\text{out}}^* + \boldsymbol{B}_{\text{out}}\boldsymbol{H}_{\text{out}}^* - \frac{1}{2}\overline{\overline{\boldsymbol{I}}}(\boldsymbol{E}_{\text{out}}^* \cdot \boldsymbol{D}_{\text{out}} + \boldsymbol{H}_{\text{out}}^* \cdot \boldsymbol{B}_{\text{out}})] \quad (1)$$

Where 'out' represents the exterior total field of the scatterer; $\boldsymbol{E}$, $\boldsymbol{D}$, $\boldsymbol{H}$ and $\boldsymbol{B}$ are the electric field, displacement vector, magnetic field and induction vectors respectively, $\langle \rangle$ represents the time average and $\overline{\overline{\boldsymbol{I}}}$ is the unity tensor.

On the other hand, based on the Lorentz force, the total force (surface force and the bulk force [40-42]) can be written as:

$$\left\langle \boldsymbol{F}_{\text{Total}} \right\rangle = \left\langle \boldsymbol{F}_{\text{Volume}} \right\rangle = \left\langle \boldsymbol{F}_{\text{Bulk}} \right\rangle + \left\langle \boldsymbol{F}_{\text{Surf}} \right\rangle = \int \left\langle \boldsymbol{f}_{\text{Bulk}} \right\rangle dv + \int \left\langle \boldsymbol{f}_{\text{Surface}} \right\rangle ds \quad (2)$$

Where

$$\left\langle \boldsymbol{f}_{\text{Surface}} \right\rangle = [\sigma_e \boldsymbol{E}_{avg}^* + \sigma_m \boldsymbol{H}_{avg}^*]_{\text{at } r=a}$$
$$= \{\epsilon_b (\boldsymbol{E}_{\text{out}} - \boldsymbol{E}_{\text{in}}) \cdot \hat{\boldsymbol{n}}\} \left( \frac{\boldsymbol{E}_{\text{out}} + \boldsymbol{E}_{\text{in}}}{2} \right)^*_{\text{at } r=a} + \{\mu_b (\boldsymbol{H}_{\text{out}} - \boldsymbol{H}_{\text{in}}) \cdot \hat{\boldsymbol{n}}\} \left( \frac{\boldsymbol{H}_{\text{out}} + \boldsymbol{H}_{\text{in}}}{2} \right)^*_{\text{at } r=a}, \quad (3)$$

$\boldsymbol{f}_{\text{Surface}}$ is the surface force density (the force which is felt by the bound electric and magnetic surface charges of a scatterer), which is calculated just at the boundary of a scatterer [40-42]. 'in' represents the interior fields of the scatterer; 'avg' represents the average of the field. $\sigma_e$ and $\sigma_m$ are the bound electric and magnetic surface charge densities of the scatterer respectively. The unit vector $\hat{n}$ is an outward pointing normal to the surface. $\varepsilon_b$ is permittivity and $\mu_b$ is permeability of the background.

$$\langle \boldsymbol{f}_{\text{Bulk}} \rangle = \frac{1}{2}\text{Re}[\varepsilon_0 (\nabla \cdot \boldsymbol{E}_{\text{in}})\boldsymbol{E}_{\text{in}}^* + \mu_0 (\nabla \cdot \boldsymbol{H}_{\text{in}})\boldsymbol{H}_{\text{in}}^*] - \frac{1}{2}\text{Re}[i\omega(\varepsilon_s - \varepsilon_b)\{\boldsymbol{E}_{\text{in}} \times \boldsymbol{B}_{\text{in}}^*\} + i\omega(\mu_s - \mu_b)\{\boldsymbol{D}_{\text{in}}^* \times \boldsymbol{H}_{\text{in}}\}] \quad (4)$$

$\boldsymbol{f}_{\text{Bulk}}$ is the bulk force density, which is calculated from the interior of the scatterer by employing the inside field [40-42]. $\varepsilon_s$ is permittivity and $\mu_s$ is permeability of the scatterer. As per we know, the Lorentz force dynamics of plasmonic particles and specially dimers have not been discussed previously. It is notable that the 'external dipolar force' [43-48] (which has also been described as Lorentz force in [43]) is quite different than the Lorentz force [40-42, 49, 50] defined in our Eqs (2)-(4).

**ACKNOWLEDGEMENTS**

M.R.C.M. acknowledges Associate Professor Qiu Cheng Wei in National University of Singapore (NUS) for important discussions throughout this work. M.R.C.M. also acknowledges Mei Shengtao (PhD student in NUS) and Dihan Hasan (PhD student in NUS) for some initial discussions regarding Lumerical software. W.D. acknowledges National Natural Science Foundation of China under grant number 11474077.

**FIGURE AND CAPTION LIST**

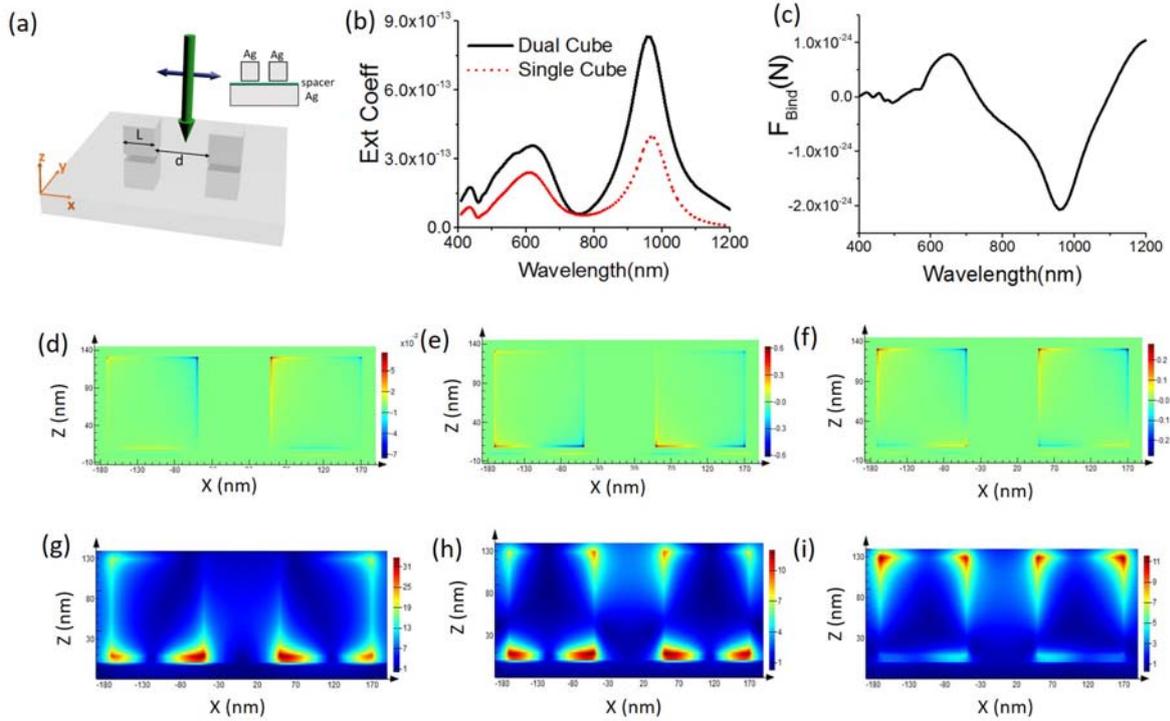

Figure 1. (a) Two silver cubes (L=120 nm) are placed over silver substrate and their inter-particle gap, *d*, is 100 nm (spacer height 5 nm from the bare substrate and the cubes are placed 5 nm away from the spacer). (b) Comparison of the extinction coefficients of two silver cubes [set-up of Fig. (a)] and a single silver cube placed over silver substrate. (c) The binding force of set-up (a). Our calculated force value is in the range of $10^{-24}$ Newton, as the incident intensity is 0.001176 W/m$^2$. On the other hand, in [27] the incident intensity is $0.05\times10^{12}$ W/m$^2$ and their force is in Pico-Newton range. So, force in our work would also appear in Pico-Newton range ($10^{-24}/0.001176\times0.05\times10^{12}\sim42.52$ pN), if the incident intensity in our work were $0.05\times10^{12}$ W/m$^2$ instead of 0.001176 W/m$^2$. (d)-(f): charge distribution at wavelengths: 966 (DD), 754 (Fano) and 627 (QQ) nm respectively. D and Q represent dipole and quadrupole respectively. (g)-(i): Electric field distribution for those same wavelengths respectively.

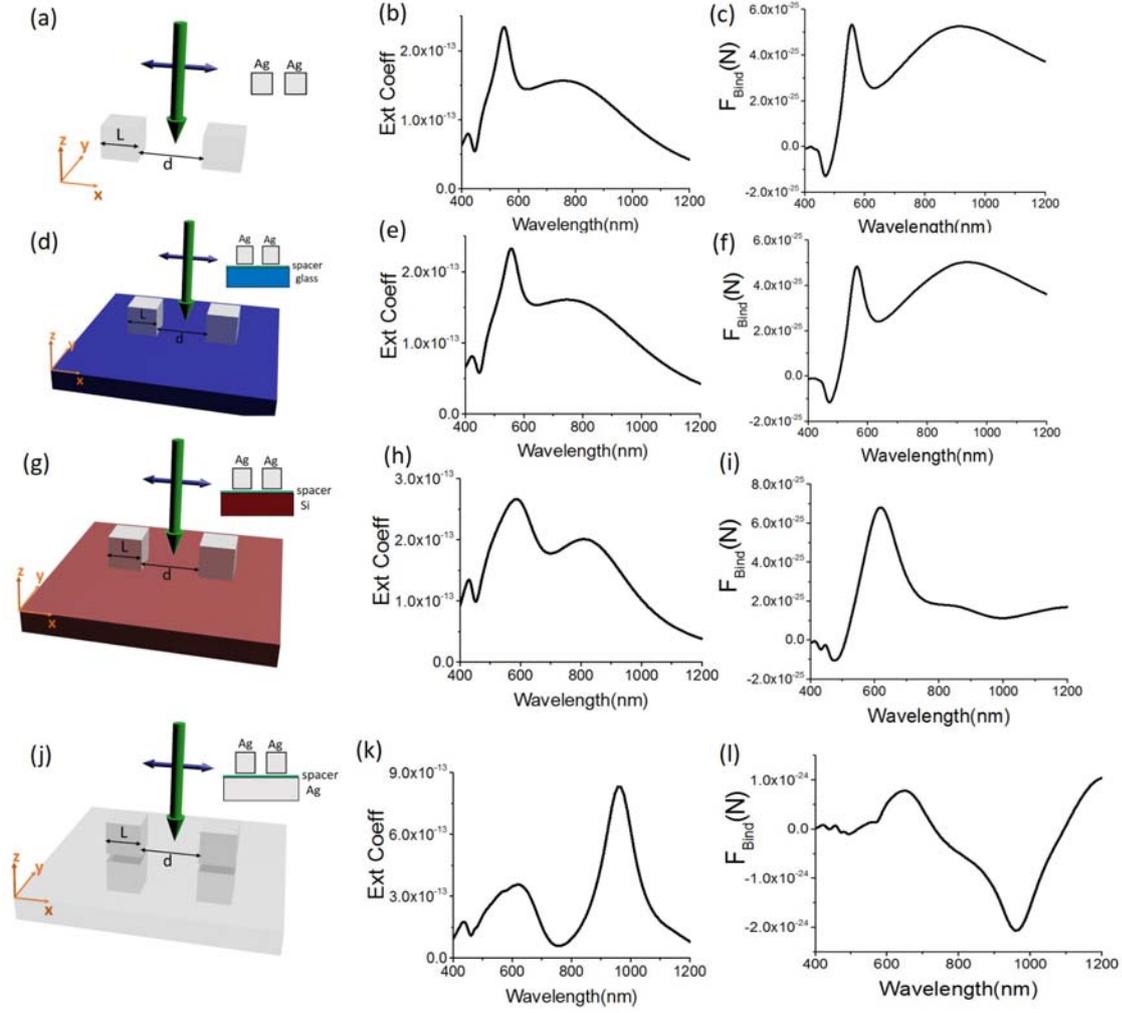

Figure 2. Two silver cubes (L=120 nm) are placed over different substrates and their inter-particle gap, $d$, is 150 nm. The spacer height is always 5nm. Cubes are placed 5nm (h= 5nm) away from the spacer. $x$-polarized plane wave is propagating towards $-z$-direction. (a)-(c): No substrate is placed; the extinction coefficient and binding force for that configuration. (d)-(f): Glass substrate (refractive index 1.5) is placed; the extinction coefficient and binding force for that configuration. (g)-(i): Silicon substrate is placed; the extinction coefficient and binding force for that configuration. (j)-(l): Silver substrate is placed; the extinction coefficient and binding force for that configuration. It should be noted that our calculated force value is in the range of $10^{-24}$ Newton, as the incident intensity is 0.001176 W/m$^2$ (with the incident amplitude of 1V/m). On the other hand, in [27] the incident intensity is $0.05\times10^{12}$ W/m$^2$ and their force is in Pico-Newton range. So, force in our work would also appear in Pico-Newton range, if the incident intensity in our work were $0.05\times10^{12}$ W/m$^2$ [27] instead of 0.001176 W/m$^2$.

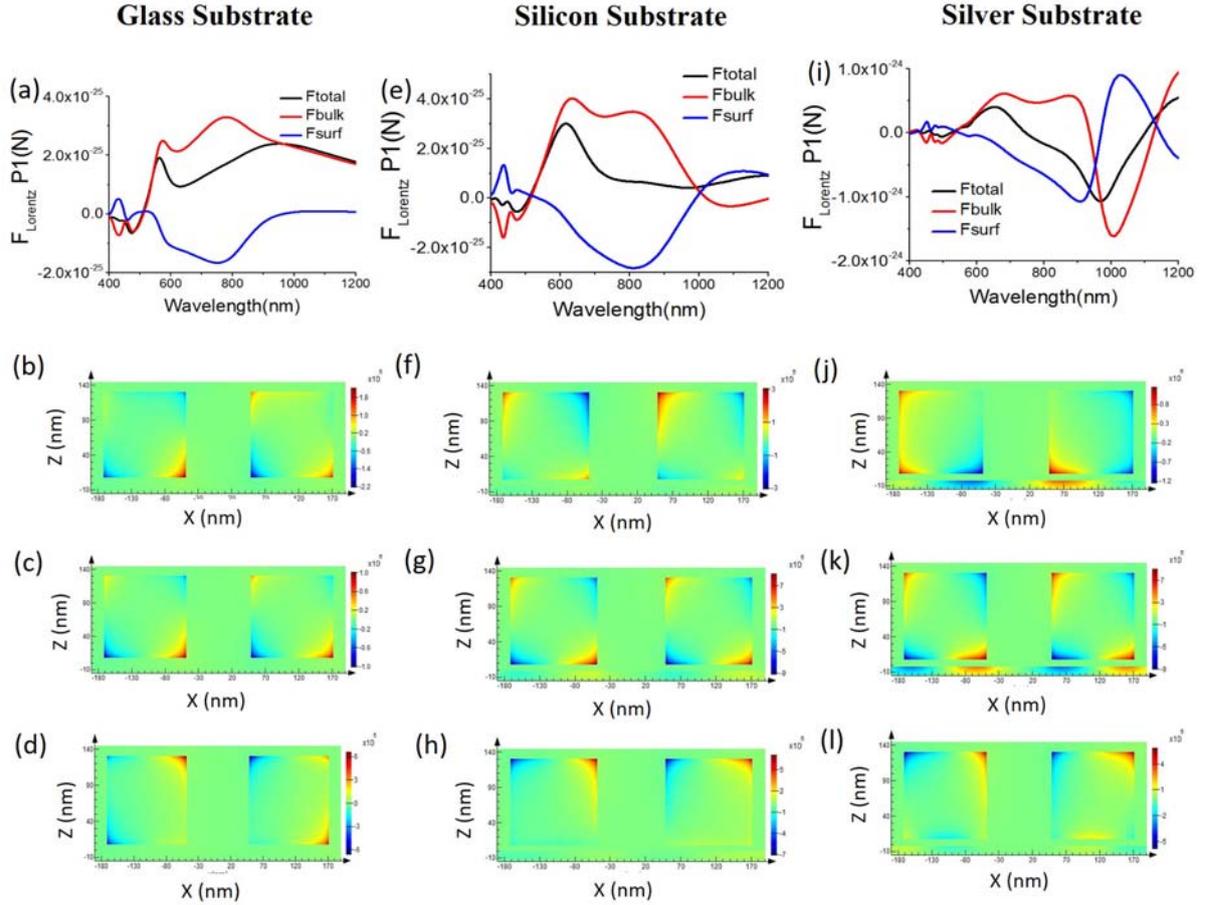

Figure 3. For the cases of different substrates, Lorentz force components and the steady state current ($J_y$) of two silver cubes along with the substrate. First row represents the Lorentz force components: surface force, bulk force [cf. Eq. (3) and (4) in 'METHOD' section] and total force [cf. Eq. (1) and (2) in 'METHOD' section] only on cube-1 (left cube, denoted as P1). 'o' represents the chosen wavelengths for which the steady state current distributions are plotted later. Second, third and fourth row represent the steady state current ($J_y$) from front view [in xz plane; setting the window very close to the cube surfaces from front view] for three different wavelengths (marked as 'o') respectively: (a)-(d): For glass substrate where wavelengths are chosen: 754, 622 and 550 nm respectively. (e)-(h): For Silicon substrate where resonance wavelengths are: 816 (DD), 679 (Fano) and 578 nm (QQ) respectively. (i)-(l): For Silver substrate where resonance wavelengths are: 966 (DD), 754 (Fano) and 627 nm (QQ) respectively.

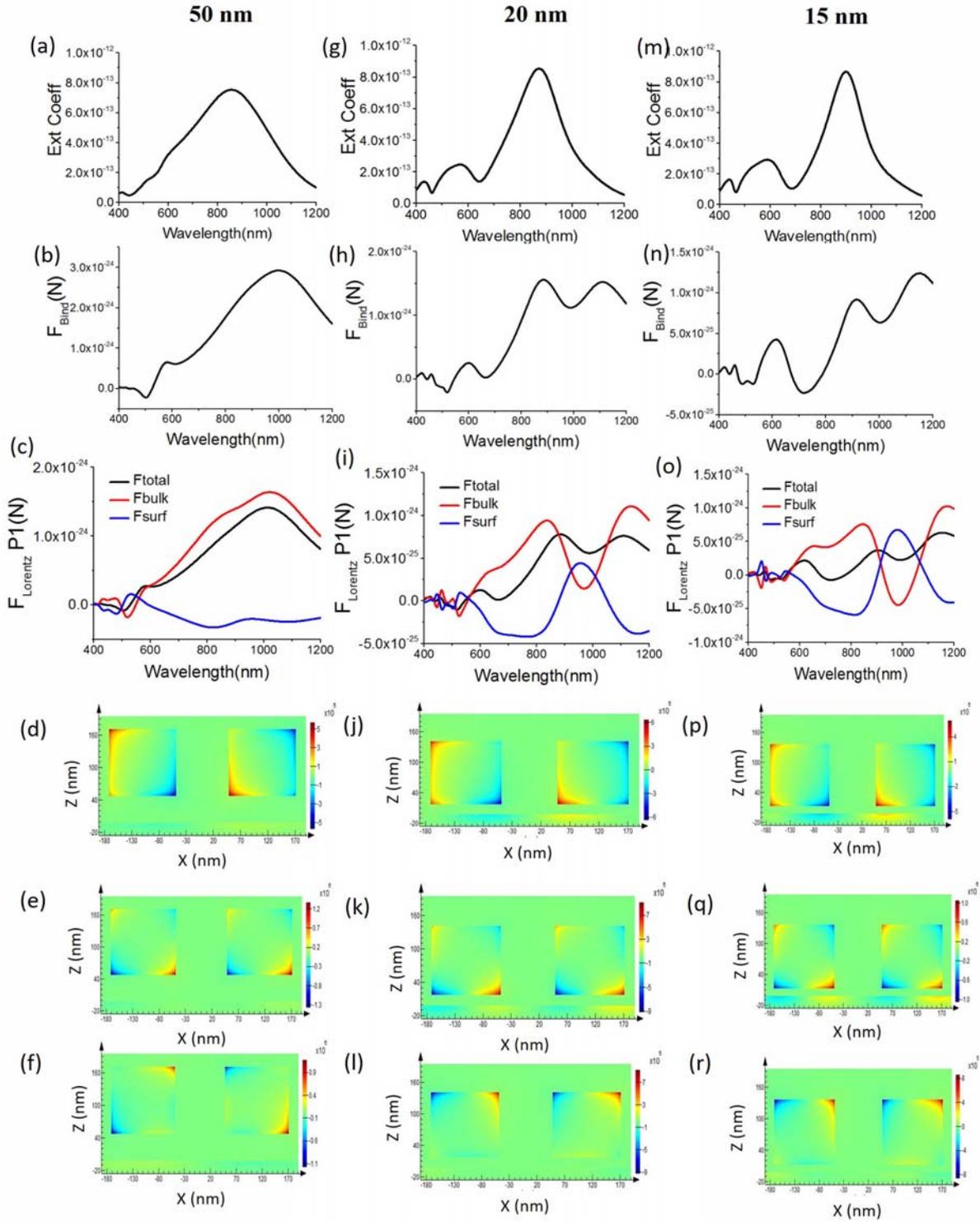

Figure 4. Two silver cubes (L=120 nm) are placed at different heights from the silver substrates (three different columns represent the position of the cubes: 50, 20 and 15 nm away from the substrate) and their inter-particle gap, $d$, is fixed 100 nm. The spacer height is 5nm. $x$-polarized plane wave is propagating towards $-z$-direction. For different heights from the substrate: First row: extinction co-efficient; second row: binding force; 'o' represents the chosen wavelengths for which the steady state current distributions are plotted later. Third row: Lorentz force components along with the total force only on cube-1 (left cube, denoted as P1); fourth, fifth and sixth row: steady state current ($J_y$) from front view [in xz plane; setting the window very close to the cube surfaces from front view] for two silver cubes along with the substrate in three different wavelengths (marked as 'o'). For first column [50 nm height, (d)-(f)] wavelengths are chosen as: 855, 553 and 503 nm respectively. For second column [20 nm height, (j)-(l)] resonance wavelengths are: 872 (DD), 631 (Fano) and 567 nm (QQ) respectively. For third column [15 nm height, (p)-(r)] resonance wavelengths are: 890 (DD), 674 (Fano) and 590 nm (QQ) respectively.

# REFERENCES

[1]  U. Fano, Effects of Configuration Interaction on Intensities and Phase Shifts. Phys. Rev. 124, 1866 (1961).

[2] Y. S. Joe, A. M. Satanin, and C. S. Kim, Classical analogy of Fano resonances. Phys. Scr. 74, 259 (2006).

[3] A. E. Miroshnichenko, S. Flach, and Y. S. Kivshar, Fano resonances in nanoscale structures. Rev. Mod. Phys. 82, 2257 (2010).

[4] L. V. Brown, H. Sobhani, J. B. Lassiter, P. Nordlander, and N. J. Halas, Heterodimers: plasmonic properties of mismatched nanoparticle pairs. Acs Nano 4, 819 (2010).

[5] O. Peña-Rodríguez, U. Pal, M. Campoy-Quiles, L. Rodríguez-Fernández, M. Garriga,  and M. I. Alonso, Enhanced Fano resonance in asymmetrical Au: Ag heterodimers. The Journal of Physical Chemistry C 115, 6410 (2011).

[6]  S. S. Acimovic, M. P. Kreuzer, M. U. Gonzalez, R. Quidant, Plasmon near-field coupling in metal dimers as a step toward single-molecule sensing. ACS Nano 3, 1231(2009).

[7]  N. Verellen, P. Van Dorpe, C. J. Huang, K. Lodewijks, G. A. E. Vandenbosch, L. Lagae,   and V. V. Moshchalkov, Plasmon Line Shaping Using Nanocrosses for High Sensitivity Localized Surface Plasmon Resonance Sensing. Nano Lett. 11, 391 (2011).

[8] A. M. Michaels, J. Jiang, L. Brus, Ag nanocrystal junctions as the site for surface-enhanced Raman scattering of single rhodamine 6G molecules. J. Phys. Chem. B 104, 11965 (2000).

[9]  K. R. Li, M. I. Stockman, D. J. Bergman, Self-similar chain of metal nanospheres as an efficient nanolens. Phys. Rev. Lett. 91, 227402 (2003).

[10] S. Zhang, K. Bao, N. J. Halas, H. Xu, and P. Nordlander, Substrate-induced Fano resonances of a plasmonic nanocube: a route to increased-sensitivity localized surface plasmon resonance sensors revealed. Nano Letters 11, 1657 (2011).

[11] T. Shegai, S. Chen, V. D. Miljkovic, G. Zengin, P. Johansson, Käll, and M., A Bimetallic Nanoantenna for Directional Colour Routing. Nat. Comm. 2, 481(2011).

[12] Q. Zhang, J. J. Xiao, X. M. Zhang, Y. Yao, and H. Liu, Reversal of optical binding force by Fano resonance in plasmonic nanorod heterodimer. Opt. Express 21, 6601 (2013).

[13] Q. Zhang and J. J. Xiao, Multiple reversals of optical binding force in plasmonic disk-ring nanostructures with dipole-multipole Fano resonances. Optics letters 38, 4240 (2013).

[14] S. Zhang, K. Bao, N. J. Halas, H. Xu, and P. Nordlander, Substrate-induced Fano resonances of a plasmonic nanocube: a route to increased-sensitivity localized surface plasmon resonance sensors revealed. Nano Letters 11, 1657 (2011).

[15] D. Sikdar, W. Zhu, W. Cheng, and M. Premaratne, Substrate-Mediated Broadband Tunability in Plasmonic Resonances of Metal Nanoantennas on Finite High-Permittivity Dielectric Substrate. Plasmonics 10, 1663(2015).

[16] M. W. Knight, Y. Wu, J. B. Lassiter, P. Nordlander, and N. J. Halas, Substrates matter: influence of an adjacent dielectric on an individual plasmonic nanoparticle. Nano Letters 9, 2188 (2009).

[17] K. Joulain, R. Carminati, J. P. Mulet, and J. J. Greffet, Definition and measurement of the local density of electromagnetic states close to an interface. Phys. Rev. B 68, 245405 (2003).

[18] Z. Wu, and Y. Zheng, Radiative Enhancement of Plasmonic Nanopatch Antennas. Plasmonics 11, 213 (2016).

[19] W. R. Holland, and D. G. Hall, Frequency shifts of an electric-dipole resonance near a conducting surface. Phys. Rev. Lett. 52, 1041(1984).

[20] O. Brzobohatý, T. Čižmár, V. Karásek, M. Šiler, K. Dholakia, and P. Zemánek, Experimental and theoretical determination of optical binding forces. Opt. Express 18, 25389 (2010).

[21] J. B. Lassiter, F. McGuire, J. J. Mock, C. Ciracì, R. T. Hill, B. J. Wiley, and D. R. Smith, Plasmonic waveguide modes of film-coupled metallic nanocubes. Nano Letters 13, 5866 (2013).

[22] C. Manolatou and F. Rana, Subwavelength nanopatch cavities for semiconductor plasmon lasers. IEEE Journal of Quantum Electronics 44, 435(2008).

[23] A. Moreau, C. Ciracì, J. J. Mock, R. T. Hill, Q. Wang, B. J. Wiley, and D. R. Smith, Controlled-reflectance surfaces with film-coupled colloidal nanoantennas. Nature 492, 86 (2012).

[24] A. J. Hallock, P. L. Redmond, and L. E. Brus, Optical forces between metallic particles. Proc Natl Acad Sci 102, 1280 (2005).

[25] P. Chu and D. L. Mills, Laser-induced forces in metallic nanosystems: The role of plasmon resonances. Phys. Rev. Lett. 99, 127401(2007).

[26] Z. Xiao-Ming, X. Jun-Jun, and Z. Qiang, Optical binding forces between plasmonic nanocubes: A numerical study based on discrete-dipole approximation. Chinese Phys. B 23, 017302 (2013).

[27] J. Ng, R. Tang, and C. T. Chan, Electrodynamics study of plasmonic bonding and anti-bonding forces in a bisphere. Phys. Revi. B 77, 195407(2008).

[28] Q. Zhang, J. J. Xiao, X. M. Zhang, and Y. Yao, Optical binding force of gold nanorod dimers coupled to a metallic slab. Optics Communications 301, 121 (2013).

[29] H. Liu, J. Ng, S. B. Wang, Z. H. Hang, C. T. Chan, and S. N. Zhu, Strong plasmon coupling between two gold nanospheres on a gold slab. New Journal of Phys. 13, 073040 (2011).

[30] A. S. Zelenina, R. Quidant, and M. Nieto-Vesperinas, Enhanced optical forces between coupled resonant metal nanoparticles. Opt. Lett. 32, 1156 (2007).

[31] Lumerical FDTD: https://www.lumerical.com/tcad-products/fdtd/

[32]   Edward D. Palik, Handbook of optical constants of solids. Vol. 3. Academic press, 1998.

[33] G. L. Úvŕque and O. J. Martin, Optical interactions in a plasmonic particle coupled to a metallic film. Opt. Express 14, 9971(2006).

[34] R. Quidant, C. Girard, Surface-plasmon-based optical manipulation. Laser Photonics Rev. 2, 47(2008).

[35] C. Min, Z. Shen, J. Shen, Y. Zhang, H. Fang, G. Yuan, L. Du, S. Zhu, T. Lei, and X. Yuan, Focused plasmonic trapping of metallic particles. Nat. comm. 4, 2891 (2013).

[36] T. V. Raziman and O. J. Martin, Internal optical forces in plasmonic nanostructures. Opt. Express 23, 20143 (2015).

[37] P. T. Bowen and D. R. Smith, Coupled-mode theory for film-coupled plasmonic nanocubes. Phys. Rev. B 90, 195402 (2014).

[38] C.-W. Qiu, W. Ding, M.R.C. Mahdy, D. Gao, T. Zhang, F.C. Cheong, A. Dogariu, Z. Wang, and C.T. Lim, Photon momentum transfer in inhomogeneous dielectric mixtures and induced tractor beams. Light: Science and Applications 4, e278 (2015).

[39] T. Zhu, M. R. C. Mahdy, Y. Cao, H. Lv, F. Sun, Z. Jiang, and W. Ding, Optical pulling using evanescent mode in sub-wavelength channels. Opt. Express. 24, 18436 (2016).

[40] M. Li. H. Wang, D. Gao, L. Gao, J. Xu, and C. W. Qiu, Radiation pressure of active dispersive chiral slabs. Opt. Express. 23, 16546 (2015).

[41] M. Dienerowitz, M. Mazilu, and K. Dholakia, Optical manipulation of nanoparticles: a review. J. Nanophoton. 2, 021875 (2008) and B. A. Kemp, T. M. Grzegorczyk, and J. A. Kong, Lorentz Force on Dielectric and Magnetic Particles. Journal of Electromag. Wave Appl. 20, 827 (2006).

[42] Volumetric Technique of Lumerical Software:

https://kb.lumerical.com/en/nanophotonic_applications_optical_tweezers_volumetric_technique.html
and
https://kb.lumerical.com/en/index.html?nanophotonic_applications_optical_tweezers_volumetric_technique.html

[43] V. D. Miljkovic, T. Pakizeh, B. Sepulveda, P. Johansson, and M. Kall, Optical Forces in Plasmonic Nanoparticle Dimers. The Journal of Physical Chemistry C 114, 7472 (2010).

[44] M. M. Rahman, A. A. Sayem, M. R. C. Mahdy, M. E. Haque, R. Islam, S. T. R. Chowdhury, and M. A. Matin, Tractor beam for fully immersed multiple objects: Long distance pulling, trapping, and rotation with a single optical set-up. Annalen der Physik 527, 777 (2015).

[45] D. Gao, A. Novitsky, T. Zhang, F. C. Cheong, L. Gao, C. T. Lim, B. Luk'yanchuk, and C.-W. Qiu, Unveiling the correlation between non-diffracting tractor beam and its singularity in Poynting vector. Laser Photon. Rev 9, 75 (2014).

[46] A. Novitsky, C.-W. Qiu, and A. Lavrinenko, Material-independent and size-independent tractor beams for dipole objects. Phys. Rev. Lett. 109, 023902 (2012).

[47] Dongliang Gao, Weiqiang Ding, Manuel Nieto-Vesperinas, Xumin Ding, Mahdy Rahman, Tianhang Zhang, Chwee Teck Lim and Cheng-Wei Qiu, Optical Manipulation from Microscale to Nanoscale: Fundamentals, Advances, and Prospects. Light: Science and Applications 6, e17039 (2017).

[48] Tianhang Zhang, Mahdy Rahman Chowdhury Mahdy, Yongmin Liu, Jing hua Teng, Lim Chwee Teck, Zheng Wang and Cheng-Wei Qiu, All-Optical Chirality-sensitive Sorting via Reversible Lateral Forces in Interference Fields. ACS Nano (2017).

[49] M.R.C. Mahdy, Tianhang Zhang, Weiqiang Ding, Amin Kianinejad, Manuel Nieto-Vesperinas, Consistency of time averaged optical force laws for embedded chiral and achiral objects. *arXiv:1704.00334* (2017).

[50] M.R.C. Mahdy , M. Q. Mehmood, W. Ding, T. Zhang and Z. N. Chen, Lorentz force and the optical pulling of multiple rayleigh particles outside the dielectric cylindrical waveguides. Annalen der Physik 529 (2016).